\documentclass[a4paper]{jpconf}
\usepackage{graphicx}
\usepackage{epstopdf}
\usepackage{amsmath}
\usepackage{multirow}
\usepackage{array}

\begin{document}
\title{Antibaryon--nucleus bound states}

\author{J. Hrt\'{a}nkov\'{a}$^{1,2}$, J. Mare\v{s}$^1$}

\address{$^1$ Nuclear Physics Institute, 25068 \v{R}e\v{z}, Czech Republic \\ 
$^2$ Czech Technical University in Prague, Faculty of Nuclear Sciences and Physical 
Engineering, B\v{r}ehov\'{a} 7, 115 19 Prague 1, Czech Republic }

\ead{hrtankova@ujf.cas.cz}

\begin{abstract}
We calculated antibaryon ($\bar{B}$ = $\bar{p}$, $\bar{\Lambda}$, $\bar{\Sigma}$, $\bar{\Xi}$) 
bound states in selected nuclei within the relativistic mean-field (RMF) model. The G-parity motivated 
$\bar{B}$--meson coupling constants were scaled to yield corresponding potentials consistent with available 
experimental data. Large polarization of the nuclear core caused by $\bar{B}$ was confirmed. 
The $\bar{p}$ annihilation in the nuclear medium was incorporated by including a phenomenological
imaginary part of the optical potential. The calculations using a complex $\bar{p}$--nucleus potential 
were performed fully self-consistently. The $\bar{p}$ widths significantly decrease when the phase 
space reduction is considered for $\bar{p}$ annihilation products, but they still remain sizeable
 for potentials consistent with $\bar{p}$--atom data.
\end{abstract}

\section{Introduction}
The study of antibaryon--nucleus interactions has attracted increasing interest in recent years at the prospect 
of future experiments at the FAIR facility \cite{FAIR}. In particular, much attention has been devoted to the 
antiproton--nucleus interaction and the possibility of formation of $\bar{p}$--nucleus bound states 
\cite{Mishustin, lari mish satarov, lenske}. Exploring the $\bar{p}$--nucleus interaction could provide 
valuable information about the behavior of the antiproton in the nuclear medium as well as nuclear dynamics. 
One of the motivations for our study of $\bar{p}$--nucleus bound states is the conjecture that the 
considerable suppression of the phase space for the $\bar{p}$ annihilation products in the nuclear
 medium could lead to relatively long living $\bar{p}$ inside the nucleus \cite{Mishustin}. 

In this contribution, we report on our recent fully self-consistent calculations of antibaryon--nucleus 
bound states within the relativistic mean-field model \cite{Walecka}. The behavior of an antibaryon in 
the nuclear medium and the dynamical effects caused by the presence of the antibaryon in the nucleus were 
studied for several selected nuclei. Special attention was devoted to the $\bar{p}$--nucleus interaction 
including $\bar{p}$ absorption in the nucleus. 

In Section 2, a brief description of the underlying model 
is given. Few representative results of our calculations are presented and discussed in Section 3. 

\section{Model}
In the present work, antibaryon--nucleus bound states are studied within the framework of the RMF approach 
applied to a system of nucleons and one antibaryon ($\bar{B}$ = $\bar{p}$, $\bar{\Lambda}$, $\bar{\Sigma}$, $\bar{\Xi}$). The interaction among (anti)baryons is mediated by the exchange of the scalar ($\sigma$)
 and vector ($\omega_{\mu}$, $\vec{\rho}_\mu$) meson fields, and the massless photon field $A_{\mu}$. The standard Lagrangian density $\mathcal{L}_N$ for nucleonic 
sector is extended by the Lagrangian density $\mathcal{L}_{\bar{B}}$ describing the antibaryon 
interaction with the nuclear medium (see ref.~\cite{jarka} for details). The variational principle yields the equations of motion 
for the hadron fields involved. The Dirac equations for nucleons and antibaryon read:  
\begin{equation} \label{Dirac antiproton}
[-i\vec{\alpha}\vec{\nabla} +\beta(m_j + S_j) + V_j]\psi_j^{\alpha}=\epsilon_j^{\alpha} \psi_j^{\alpha}, 
\quad j=N,\bar{B}~,
\end{equation}
where
\begin{equation}
S_j=g_{\sigma j}\sigma, \quad V_j=g_{\omega j} \omega_0 + g_{\rho j}\rho_0 \tau_3 + e_j \frac{1+\tau_3}{2}A_0
\end{equation}
are the scalar and vector potentials. Here, $\alpha$ denotes single particle states, $m_j$ stands for 
(anti)baryon masses and $g_{\sigma j}, g_{\omega j}, g_{\rho j}$, and $e_j$ are (anti)baryon coupling 
constants to corresponding fields. The presence of $\bar{B}$ induces additional source terms in the Klein--Gordon equations for the meson fields:
\begin{equation}
\begin{split} \label{meson eq}
(-\triangle + m_\sigma^2+ g_2\sigma + g_3\sigma^2)\sigma&=- g_{\sigma N} \rho_{SN}-g_{\sigma \bar{B}} 
\rho_{S \bar{B}} \\
(-\triangle + m_\omega^2 +d\omega^2_0)\omega_0&= g_{\omega N}\rho_{VN} +g_{\omega \bar{B}} \rho_{V\bar{B}}\\
(-\triangle + m_\rho^2)\rho_0&= g_{\rho N}\rho_{IN} +g_{\rho \bar{B}}\rho_{I \bar{B}}\\
-\triangle A_0&= e_N \rho_{QN}+e_{\bar{B}}\rho_{Q\bar{B}}~,
\end{split}
\end{equation}
where $\rho_{\text{S}j}, \rho_{\text{V}j}, \rho_{\text{I}j}$ and $\rho_{\text{Q}j}$  are the scalar, 
vector, isovector and charge densities, respectively, and $m_\sigma, m_\omega, m_\rho$ are the masses of
considered mesons. The system of coupled Dirac  \eqref{Dirac antiproton}  and Klein--Gordon  
\eqref{meson eq} equations represents a self-consistent problem which is to be solved by iterative procedure.

The values of the nucleon--meson coupling constants and meson masses were adopted from the nonlinear RMF 
model TM1(2) \cite{Toki} for heavy (light) nuclei. We used also the density--dependent model TW99 
\cite{TypelWolter, TW99} in which the couplings are a function of the baryon density. The hyperon--meson 
coupling constants for the $\omega$ and $\rho$ fields are obtained using SU(6) symmetry relations. The coupling 
constants for the $\sigma$ field are constrained by available experimental data --- $\Lambda$ hypernuclei 
\cite{hyperon couplings}, $\Sigma$ atoms \cite{sigma atomy}, and $\Xi$ production in $(K^+, K^-)$ reaction 
\cite{Khaustov}.

In the RMF approach, the nuclear ground state is well described by the attractive scalar potential 
$S\simeq -350$~MeV and by the repulsive vector potential $V\simeq 300$~MeV. The resulting central 
potential acting on a nucleon in a nucleus is then approximately $S+V\simeq -50$~MeV deep. 
The $\bar{B}$--nucleus interaction is constructed from the $B$--nucleus interaction with the help of the 
G-parity transformation: the vector potential generated by the $\omega$ meson exchange thus changes its 
sign and becomes attractive. As a consequence, the total potential acting on a $\bar{B}$ will be strongly 
attractive. In particular, the $\bar{p}$--nuclear potential would be $S-V\simeq -650$~MeV deep. However, 
the G-parity transformation should be regarded as a mere starting point to determine the $\bar{B}$--meson 
coupling constants. Various many-body effects, as well as the presence of strong annihilation channels 
could cause significant deviations from the G-parity values in the nuclear medium. Indeed, the available 
experimental data from $\bar{p}$ atoms \cite{mares} and $\bar{p}$ scattering off nuclei \cite{antiNN interaction} 
suggest that the depth of the real part of the $\bar{p}$--nucleus potential is in the range $-(100 - 300)$~MeV in the 
nuclear interior. Therefore, we introduce a scaling factor $\xi \in \langle0,1\rangle$ for the antibaryon--meson 
coupling constants which are in the following relation to the baryon--meson couplings:
\begin{equation} \label{reduced couplings}
g_{\sigma \bar{B}}=\xi\, g_{\sigma N}, \quad g_{\omega \bar{B}}=-\xi\, g_{\omega N}, \quad g_{\rho 
\bar{B}}=\xi\, g_{\rho N}~.
\end{equation}

\bigskip
The annihilation of an antibaryon inside the nuclear medium is an inseparable part of any realistic
 description of the $\bar{B}$--nucleus interaction. In our calculations, only the $\bar{p}$ absorption 
in the nucleus has been considered so far. Since the RMF model does not address directly the absorption of 
the $\bar{p}$ in the nucleus we adopted the imaginary part of the optical potential in a `$t\rho$' 
form from optical model phenomenology \cite{mares}:
\begin{equation}
2\mu {\rm Im}V_{\text{opt}}(r)=-4 \pi \left(1+ \frac{\mu}{m_N}\frac{A-1}{A} 
\right){\rm Im}b_0 \rho(r)~,
\end{equation}
where $\mu$ is the $\bar{p}$--nucleus reduced mass. While the density $\rho$ was treated as a dynamical 
quantity determined within the RMF model, the parameter $b_0$ was constrained by fits to $\bar{p}$-atomic 
data \cite{mares}. The global fits to the $\bar{p}$-atomic data give a single value for the imaginary 
part of $b_0$, Im$b_0=1.9$ fm for all nuclei considered. 

The energy available for the $\bar{p}$ annihilation in the nuclear medium is usually expressed as 
$\sqrt{s}=~m_{\bar{p}}+m_{N}-B_{\bar{p}}-B_{N}$, where $B_{\bar{p}}$ and $B_{N}$ is the $\bar{p}$ and 
nucleon binding energy, respectively. Therefore, the phase space available for the annihilation products is considerably suppressed for the deeply bound antiproton. 
\begin{figure}[t]
\centering
\includegraphics[width=0.6\textwidth]{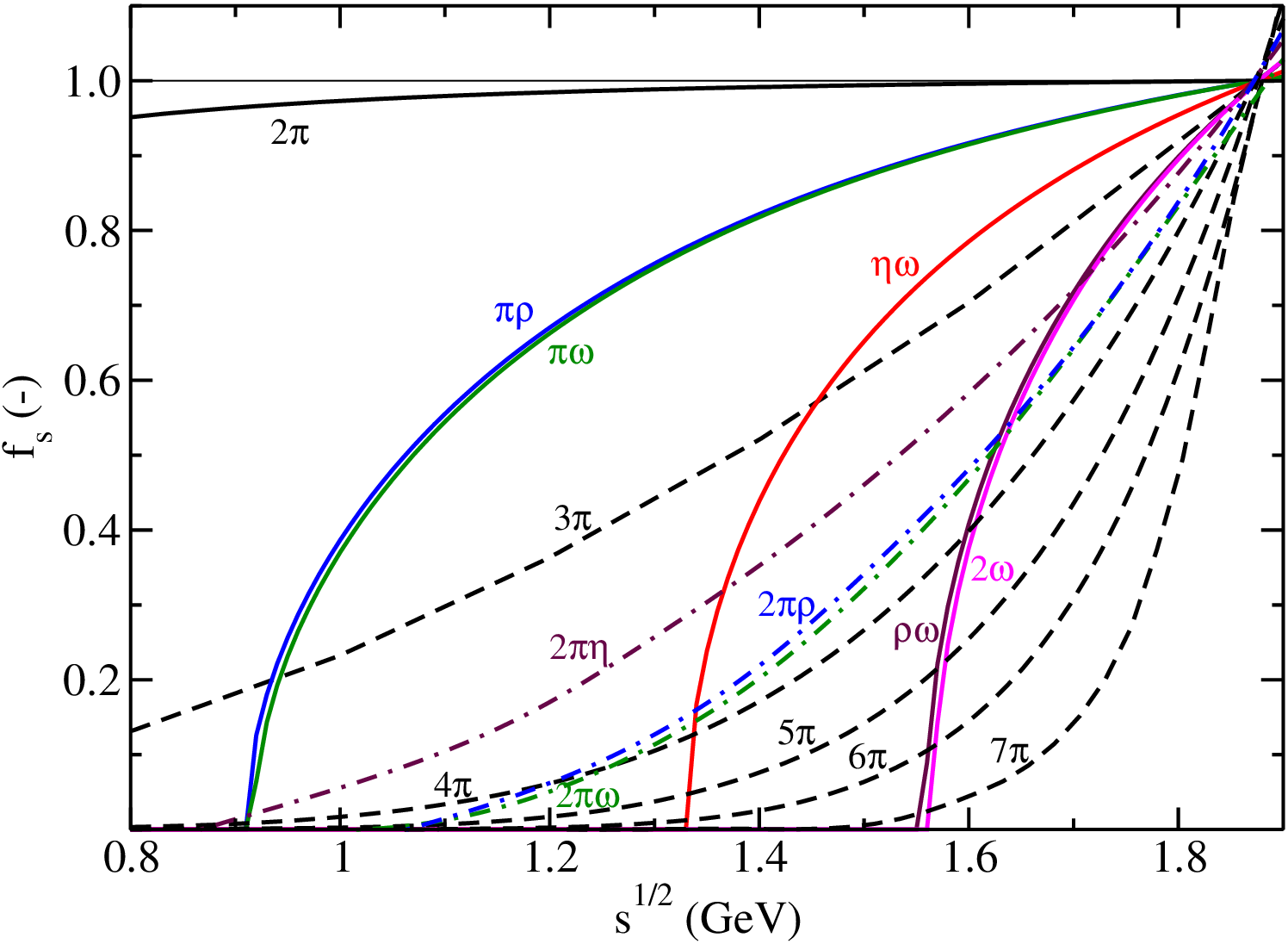}
\caption{\label{Fig.: SF}The phase space suppression factors $f_{\text{s}}$ as a function of the c.m. 
energy $\sqrt{s}$.}
\end{figure}

The phase space suppression factors ($f_{\text{s}}$) for two body decay are given by \cite{pdg}
\begin{equation}
f_{\text{s}}=\frac{M^2}{s}\sqrt{\frac{[s-(m_1+m_2)^2][s-(m_1-m_2)^2]}{[M^2-(m_1+m_2)][M^2-(m_1-m_2)^2]}}\Theta(\sqrt{s}-m_1-m_2)~,
\end{equation}
where $m_1$, $m_2$ are the masses of the annihilation products and $M=m_{\bar{p}}+m_N$.

For channels containing more than 2 particles in the final state the suppression factors $f_{\text{s}}$ were 
evaluated with the help of Monte Carlo simulation tool PLUTO \cite{PLUTO}. In Figure~\ref{Fig.: SF}, we present the phase space suppression factors 
as a function of the center-of-mass energy $\sqrt{s}$ for considered annihilation channels. 
As the energy $\sqrt{s}$ decreases many channels become significantly suppressed  
or even closed which could lead to much longer lifetime of $\bar{p}$ in a nucleus.

\section{Results}

We applied the formalism introduced in the previous section to self-consistent calculations of $\bar{p}$, 
$\bar{\Lambda}$, $\bar{\Sigma}$, $\bar{\Xi}$ bound states in selected nuclei across the periodic table. 
First, we did not consider absorption of an antibaryon in a nucleus. 
Our calculations within the TM model confirmed substantial polarization of the nuclear core caused by the 
antibaryon embedded in the nucleus~\cite{Mishustin}.  The nucleon single particle energies are significantly affected 
by the presence of $\bar{B}$ and the total binding energies increase considerably, as well. The nucleon densities in 
$\bar{p}$ nuclei reach $2-3$ times the nuclear matter density. The RMF models with constant couplings do not 
have to describe correctly the properties of nuclear matter when extrapolated to such high densities. 
Therefore, we performed calculations of $\bar{p}$ nuclei within the density--dependent model TW99, as well. 
We obtained very similar results as for the TM model and thus confirmed only small model dependence of 
our calculations.

\begin{figure}[t]
\begin{minipage}{0.5\textwidth}
\includegraphics[width=1\textwidth]{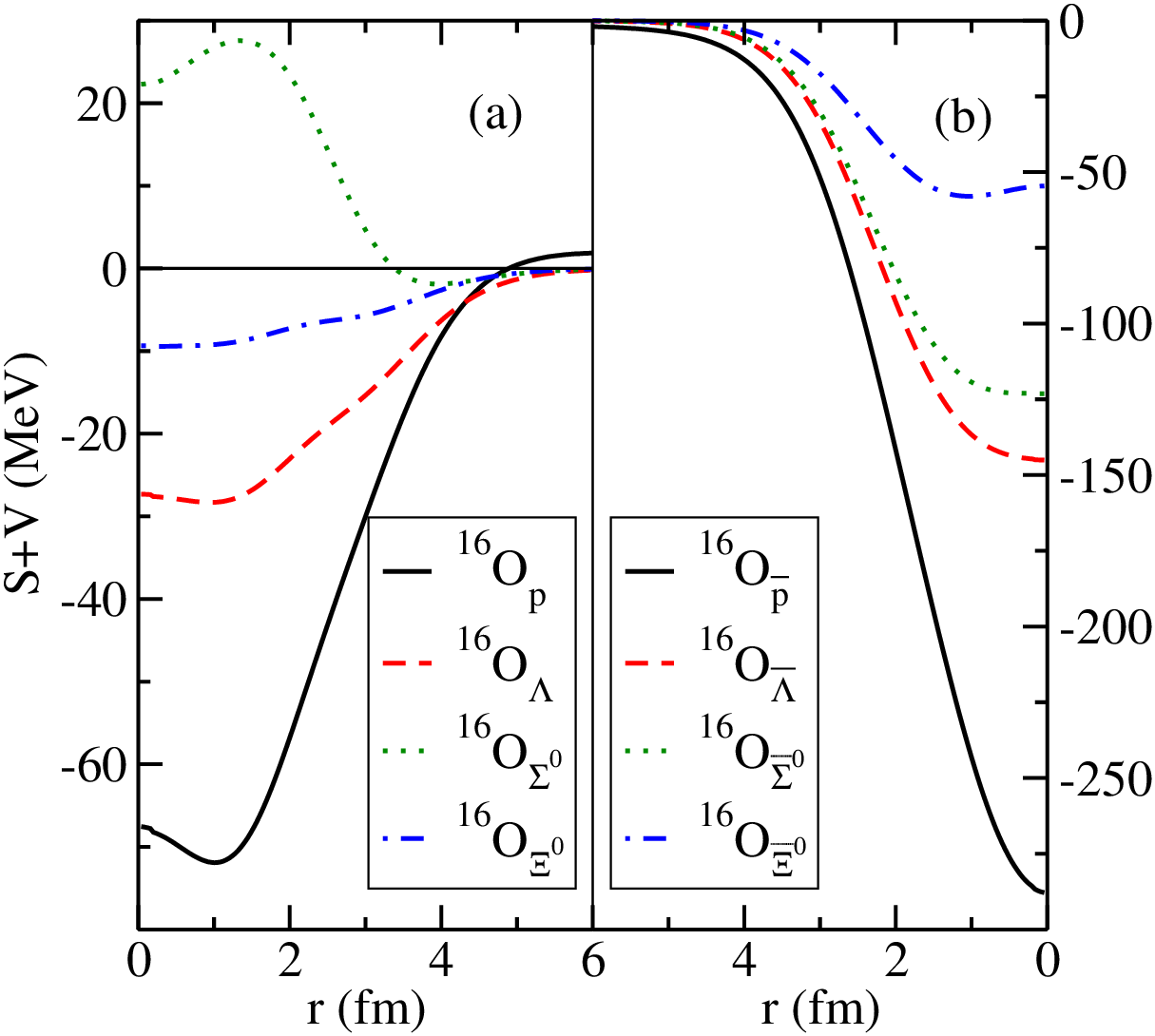}
\caption{\label{Fig.: potencialy}The $B$--nucleus (a) and $\bar{B}$--nucleus (b) potentials in $^{16}$O, calculated dynamically 
for $\xi=0.2$ in the TM2 model.}
\end{minipage}\hspace{1.2pc}%
\begin{minipage}{0.46\textwidth}
\includegraphics[width=1\textwidth]{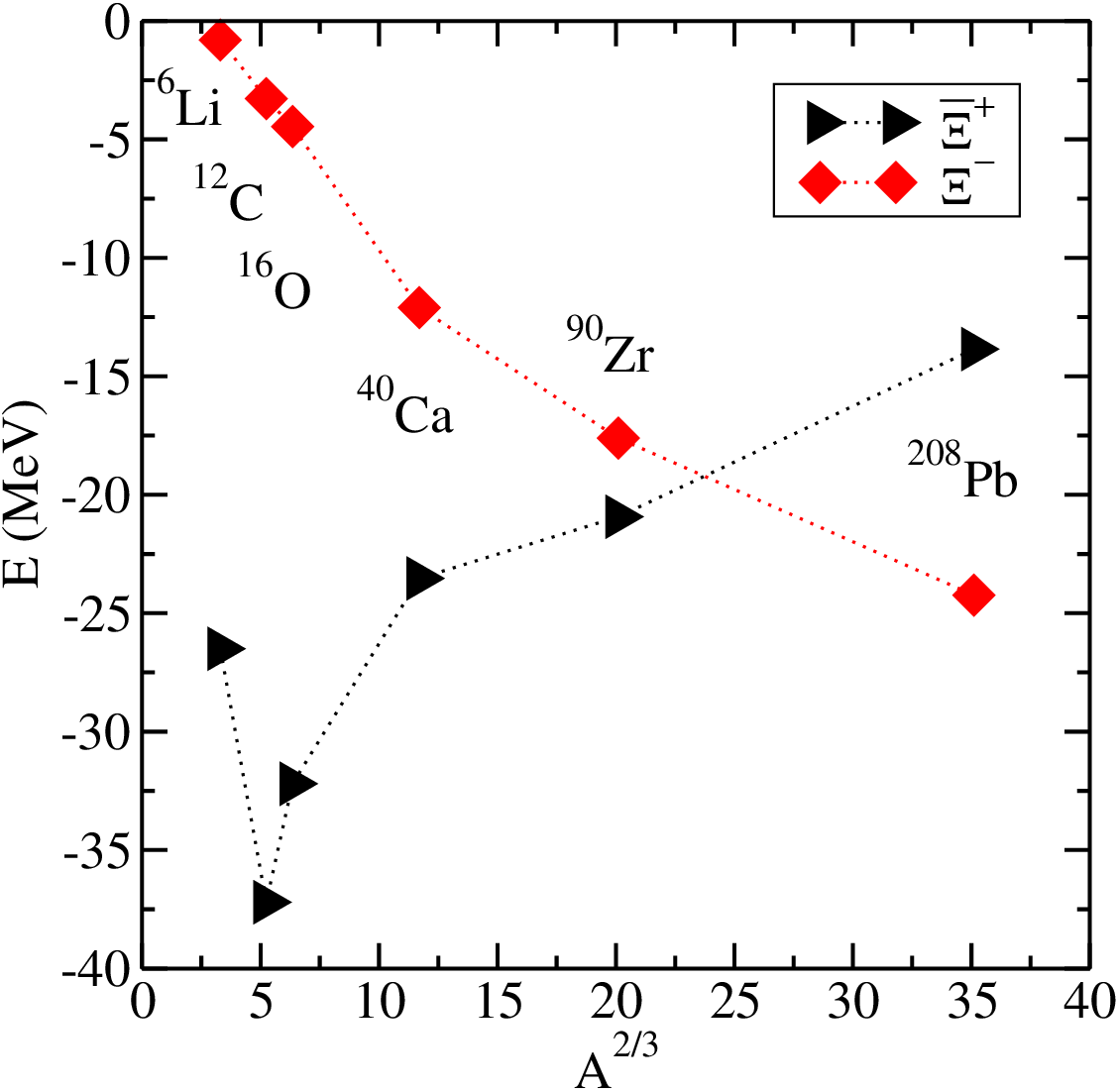}
\caption{\label{Fig.: porovnanie}Single particle energies of $\Xi^-$ and $\bar{\Xi}^+$ for $\xi=0.2$ in various nuclei, 
calculated dynamically in the TM model.}
\end{minipage}
\end{figure}

Figure \ref{Fig.: potencialy} shows the total potential acting on an extra baryon (a) and on an extra antibaryon (b) 
in $1s$ state in $^{16}$O, calculated dynamically in the TM2 model. The scaling parameter is chosen to be  $\xi = 0.2$ 
as this value gives the $\bar{p}$ potential comparable with the available experimental data.  We assume the same scaling 
parameter also for antihyperons, since there is no reliable experimental information on the in-medium antihyperon 
potentials. The potentials acting on antibaryons are fairly deep in the central region of the nucleus in contrast to the 
baryon potentials (notice that the potential for $\Sigma^0$ is even repulsive while the potential for $\bar{\Sigma}^0$ 
is strongly attractive). Such strongly attractive potentials yield deeply bound states of antibaryons in atomic nuclei.

Figure \ref{Fig.: porovnanie} presents a comparison between the $\Xi^-$ and $\bar{\Xi}^+$ $1s$ single particle energies 
in various nuclei across the periodic table, calculated dynamically in the TM model. The $\bar{\Xi}^+$ coupling constants are 
scaled by factor $\xi=0.2$. The binding energy of $\Xi^-$ is increasing with the number of nucleons in the nucleus. 
The $\bar{\Xi}^+$ binding energy follows the opposite trend and in Pb it is even less bound than $\Xi^-$. This can be 
explained by enhanced Coulomb repulsion felt by $\bar{\Xi}^+$ in heavier nuclei.

\bigskip
We performed calculations of $\bar{p}$ nuclei using a complex potential describing the $\bar{p}$ annihilation in the nuclear medium. 
The results of static as well as dynamical calculations with the real potential, complex potential, and complex potential with 
the suppression factors $f_{\text{s}}$ for $\bar{p}$ bound in $^{16}$O are presented in Table \ref{Tab.: hodnoty}. The scaling of the $\bar{p}$--meson 
coupling constants is chosen to be $\xi=0.2$. The static calculations, which do not account for the core polarization effects, give 
approximately the same values of the $\bar{p}$ single particle energy for all three cases. The single particle energies calculated 
dynamically are larger, which indicates that the polarization of the core nucleus is significant (even if the $\bar{p}$ absorption is 
taken into account). When the effect of the phase space suppression is considered the $\bar{p}$ annihilation width is substantially 
suppressed (compare $552.3$~MeV vs. $232.5$~MeV in the last row of Table \ref{Tab.: hodnoty}). However, the $\bar{p}$ width is still considerable for relevant $\bar{p}$ 
potentials consistent with the $\bar{p}$ data.

\begin{center}
\begin{table}[t]
\caption{\label{Tab.: hodnoty} The $1s$ single particle energies $E_{\bar{p}}$ and widths $\Gamma_{\bar{p}}$ (in MeV) in 
$^{16}$O$_{\bar{p}}$, calculated dynamically (Dyn) and statically (Stat) with the real, complex and complex with $f_{\text{s}}$ potentials 
(TM2 model), consistent with $\bar{p}$--atom data.}
\centering
\begin{tabular}{lcccccc}
\br
 & \multicolumn{2}{c}{Real} & \multicolumn{2}{c}{Complex} & \multicolumn{2}{c}{Complex + $f_{\text{s}}$} \\ \mr
 & Dyn & Stat & Dyn & Stat & Dyn &  Stat  \\ \mr
$E_{\bar{p}}$ & ~193.7~ & ~137.1~ & ~175.6~ & ~134.6~ & ~190.2~ & ~136.1~\\
$\Gamma_{\bar{p}}$ & ~-~ & - & 552.3 & 293.3 & 232.5 & 165.0 \\
\br
\end{tabular}
\end{table}
\end{center}

The annihilation of the $\bar{p}$ with a nucleon takes place in a nucleus. Therefore, the momentum dependent term in Mandelstam variable 
$s=(E_N + E_{\bar{p}})^2 - (\vec{p}_N + \vec{p}_{\bar{p}})^2$ is non-negligible in contrast to two body frame \cite{s}. Our self-consistent 
evaluation of $\sqrt{s}$ by considering the momenta of annihilating partners leads to an additional downward energy shift.   
As a consequence, the $\bar{p}$ width in $^{16}$O$_{\bar{p}}$ is reduced by additional $\approx 50$~MeV. We conclude that even 
after taking into account the phase space suppression corresponding to self-consistent treatment of $\sqrt{s}$ including the $\bar{p}$ and $N$ momenta, 
the $\bar{p}$ annihilation widths in nuclei remain sizeable. The corresponding lifetime of the $\bar{p}$ in the nuclear medium 
is $\simeq 1$~fm/c.

\section*{Acknowledgements}
This work was supported by GACR Grant No. P203/12/2126. We thank Pavel Tlust\'{y} for his assistance during Monte Carlo simulations using PLUTO. J. Hrt\'{a}nkov\'{a} acknowledges financial support from CTU-SGS Grant No. SGS13/216/OHK4/3T/14.

\end{document}